\documentclass[%
a4paper,
superscriptaddress,
notitlepage,
bibnotes,
nolongbibliography,
 amsmath,amssymb,
 aps,
10pt,
twocolumn,
pre,
]{revtex4-2}

\usepackage{xcolor}

\colorlet{mylinkcolor}{blue!66!black!80}

\usepackage{mathtools}

\usepackage[colorlinks=true,linkcolor=mylinkcolor,citecolor=mylinkcolor,filecolor=cyan,urlcolor=mylinkcolor,breaklinks=true]{hyperref}

\makeatletter
\@addtoreset{equation}{SIequation}
\@addtoreset{figure}{SIfigure}
\makeatother

\usepackage[utf8]{inputenc}

\newcommand{\avg}[1]{\langle#1\rangle}

\newcommand{\del}{\partial}
\newcommand{\dd}{\mathrm{d}}

\newcommand{\by}{\boldsymbol{y}}

\newcommand{\e}{{\rm e}}

\newcommand{\bra}[1]{\langle#1|}
\newcommand{\ket}[1]{|#1\rangle}
\newcommand{\kket}[1]{#1\rangle}

\begin{document}
 \title{Thermodynamic Uncertainty Relation Bounds the Extent of Anomalous Diffusion
 }
\author{David Hartich}
\email{david.hartich@mpibpc.mpg.de}
\affiliation{%
Mathematical bioPhysics Group, Max Planck Institute for Biophysical Chemistry, 37077 Göttingen, Germany}
  \author{Alja\v{z} Godec}%
  \email{agodec@mpibpc.mpg.de}
\affiliation{%
Mathematical bioPhysics Group, Max Planck Institute for Biophysical Chemistry, 37077 Göttingen, Germany}




\begin{abstract}
In a finite system driven out of equilibrium by a constant external
force 
the  thermodynamic uncertainty relation (TUR) bounds the
variance of the conjugate current variable by the thermodynamic cost of
maintaining the non-equilibrium stationary state. Here we highlight a
new facet of the TUR by showing that it also bounds the
time-scale on which a finite system can exhibit anomalous kinetics. In
particular, we demonstrate that the TUR bounds subdiffusion in a single
file confined to a ring as well as a dragged
Gaussian polymer chain even when detailed balance is
satisfied. Conversely, the TUR bounds the onset of superdiffusion
in
the active comb model. 
Remarkably, the fluctuations in a comb model evolving from a steady state behave anomalously as soon as detailed balance is broken.
Our work establishes a link between stochastic thermodynamics and the
field of anomalous dynamics that will fertilize further
investigations of thermodynamic consistency of anomalous diffusion models.
\end{abstract}

\maketitle

Imagine an overdamped random walker (e.g. a molecular motor)
moving a distance $x_t$ in a 
time $t$.
If 
driven into a non-equilibrium steady state
\cite{seif18}
the walker's mean displacement grows
linearly in time,  $\avg{x_t}=vt$ with velocity $v$, whereas the variance
$\sigma_x^2(t)\equiv \avg{x_t^2}-\avg{x_t}^2$ may exhibit anomalous
diffusion \cite{metz00,metz04,soko05,klag08,metz14} with
\begin{equation}
\sigma_x^2(t)
\simeq K_\alpha t^{\alpha}
  \label{eq:anomalous}
\end{equation}
with anomalous exponent $\alpha\neq1$ and generalized diffusion coefficient $K_\alpha$ having units $\mathrm{m}^2{\rm s}^{-\alpha}$. 
When $\alpha>1$ one speaks of superdiffusion, which was observed, for
example, in active intracellular transport \cite{casp00}, optically
controlled active media \cite{doug12}, and in evolving cell colonies
during tumor invasion \cite{malm18} to name but a few. Conversely, the
situation $\alpha<1$ is referred to as subdiffusion and  
in a biophysical context was found in observations of particles
confined to actin networks \cite{ambl96,wong04},  polymers
\cite{leg02}, denaturation bubbles in DNA \cite{hwa03}, lipid granules in yeast \cite{toli04}, and cytoplasmic RNA-proteins \cite{lamp17}.
In these systems subdiffusion is often thought to be a result of macromolecular
crowding \cite{soko12,hoef13,ghos16}, where obstacles hinder the
motion of a tracer particle.

\begin{figure}
 \centering
\includegraphics[width=\columnwidth]{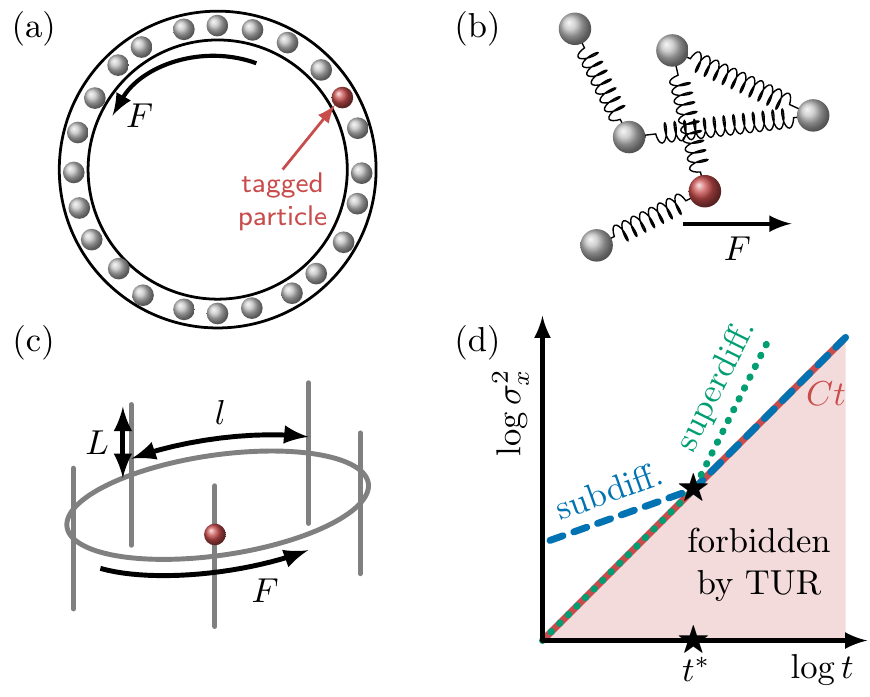}
\caption{Anomalous diffusion
  in finite
   systems. (a)~Single file on a ring driven by a force
   $F$. (b)~Tagged-particle diffusion in a harmonic chain. (c)~Biased diffusion
   in a finite (periodic) comb. The experimental observable is the unbounded displacement $x_t$
   in the direction of the force $F$.
 (d)~The TUR, $\sigma_x^2(t)\ge Ct$, delivers a threshold time $t^*$
   that imposes an upper bound on the
   duration of subdiffusion (dashed blue line) or the earliest
   possible onset of superdiffusion (dotted green line). The star
   denotes $K_\alpha (t^*)^{\alpha}=Ct^*$ in
 Eq.~\eqref{eq:tanomalous}.}
 \label{fig:anomalous_finite}
 \end{figure}

A paradigmatic example of anomalous diffusion is the motion of a
tracer particle in a single file depicted in
Fig.~\ref{fig:anomalous_finite}a where hard-core interacting particles are confined to a
one dimensional ring and block each others passage effecting the well known 
 $\alpha=1/2$ subdiffusive scaling
\cite{harr65,koll03,lin05,talo06,liza08,liza09,liza10,delf11,leib13,krap14,ryab16}
that was corroborated experimentally \cite{hahn96,wei00,lutz04}.
 Subdiffusion in single file systems emerges more generally in the
 presence of any repulsive interaction  \cite{koll03} such as, e.g. in
 polymer chains \cite{leib13,lomh14,laco15} (see
 Fig.~\ref{fig:anomalous_finite}b).
 More recently out-of-equilibrium anomalous transport was studied in
 the context of single file diffusion in the presence of a non-equilibrium bias ($v\neq0$)
 \cite{illi13,beni13,beni18,teom19} and in active
 comb models (see Fig.~\ref{fig:anomalous_finite}c) that were shown, quite surprisingly,
to display accelerated diffusion \cite{bere15} in stark contrast to passive combs (see e.g. Refs.~\cite{bouc90,bere14,beni15,sand16,lapo19}).
 
The span of anomalous diffusion in physical systems is naturally bound
to finite  (albeit potentially very long) time-scales \cite{spak19} as a result of the necessarily finite range of correlations in
a finite system that eventually ensure the emergence of the
central limit theorem \cite{hoef13}.

 We throughout consider a walker
 (e.g. a molecular motor) that operates in a (non-equilibrium) steady
 state \cite{seif18}, which means that the walker's displacement $x_t$
 is weakly ergodic. That is, the centralized displacement $x_t-vt$ 
 is unbiased with vanishing
 ``ergodicity breaking parameter'' \cite{he08} (see
 also~\cite{jeon10,cher13}), i.e. as long as trajectories are sufficiently
 long, ensemble- and time-average
 observables, such as the centralized \emph{time averaged mean square displacement} (TAMSD)
 \footnote{\unexpanded{The TAMSD is defined by $\overline{\delta^2}(t)=\lim_{\tau\to\infty}(\tau-t)^{-1}\int_\tau^t(x_{s+t}-x_{s})^2\dd s=\avg{x_t^2}
$. The centralized TAMSD is obtained by subtracting the square of the
   mean displacement along an ergodically long trajectory that reads
   $\overline{\delta}(t)=\lim_{\tau\to\infty}(\tau-t)^{-1}\int_\tau^t(x_{s+t}-x_{s})\dd
   s=\avg{x_t}=vt$.}
}, coincide.

At sufficiently long times where diffusion becomes normal,
$\sigma^2_x(t)\propto t$,  
the thermodynamic uncertainty relation (TUR) \cite{bara15,ging16}
bounds the walker's variance by \footnote{The TUR was originally proposed in the form $\epsilon^2\dot W_{\rm ss} t\ge 2 k_{\rm B}T$, where $\epsilon^2=\sigma_x^2/(vt)^2$ is the relative uncertainty and $\dot W_{\rm ss} t$ the total dissipation \cite{bara15,ging16}.}
\begin{equation}
\sigma_x^2(t) \ge \frac{2k_{\rm B}T v^2}{\dot W_{\rm ss}}t\equiv Ct,
 \label{eq:TURdef}
\end{equation}
where $\dot W_{\rm ss}$ is the power dissipated by the walker,
 $k_{\rm B} T$ is the thermal energy, and in the last step we have
defined the constant $C$. 
Eq.~\eqref{eq:TURdef} is derived by assuming that the underlying
(full) system's dynamics follows a Markovian time evolution. The TUR was originally shown to hold in the long time
limit ``$t\to\infty$'' \cite{bara15,ging16} and later on also at any
finite time for a walker's position evolving from a
non-equilibrium steady state \cite{piet17,*piet16,horo17}. 
Using aspects of information geometry \cite{ito18,dech20,ito20}
Eq.~\eqref{eq:TURdef} was recently shown to hold for any initial
condition \cite{liu20}. Subsequent 
studies have applied
Eq.~\eqref{eq:TURdef} to bound the efficiency of molecular motors
\cite{piet16b} and heat
engines \cite{shir16a,piet18}, and extended the TUR to periodically
driven systems \cite{holu18,bara18,koyu18,bara18a,koyu20}, discrete time
processes \cite{proe17}, and open quantum systems \cite{hase21}. For a broader
perspective see \cite{seif18,bara19,fala20,horo20}.

\emph{Main result.---}We now show how the TUR \eqref{eq:TURdef} may
be used to obtain a thermodynamic bound on the duration of anomalous
diffusion.
We first consider subdiffusion
($\alpha<1$) and estimate the largest time $t^*$
where  Eq.~\eqref{eq:anomalous} must cease to hold as a result of
thermodynamic consistency. 
Namely, according to \eqref{eq:TURdef} subdiffusion
in Eq.~\eqref{eq:anomalous} with
constant exponent $\alpha<1$ cannot persist beyond 
  \begin{equation}
 t^*\simeq\bigg(\frac{ K_\alpha}{C}\bigg)^{1/(1-\alpha)},
 \label{eq:tanomalous}
\end{equation}
 see intersecting point in Fig.~\ref{fig:anomalous_finite}d.
 Conversely, superdiffusion with an exponent $\alpha>1$ in
 Eq.~\eqref{eq:anomalous} cannot emerge \emph{before} $t^*$ (see
 Fig.~\ref{fig:anomalous_finite}d). Eq.~\eqref{eq:tanomalous} thus
 bounds the extent of both sub- and superdiffusion.
 The bridge between anomalous diffusion and stochastic
 thermodynamics embodied in Eq.~\eqref{eq:tanomalous} is the main
 result of this Letter. We note that the bound
 $t^*$ follows directly from the inequality \eqref{eq:TURdef} and in
 general can \emph{not} be deduced from the long time
 diffusion behavior (for an explicit counter-example see Supplemental Material (SM)  \footnote{See Supplemental Material, which includes Refs.~\cite{spec08,bark10,hart20arxiv,lapo18}, for explicit and detailed
  calculations.}).
 In the following we use the three paradigmatic physical models depicted in
 Fig.~\ref{fig:anomalous_finite} to illustrate how to apply the bound
 \eqref{eq:tanomalous}.

\begin{figure}
\includegraphics{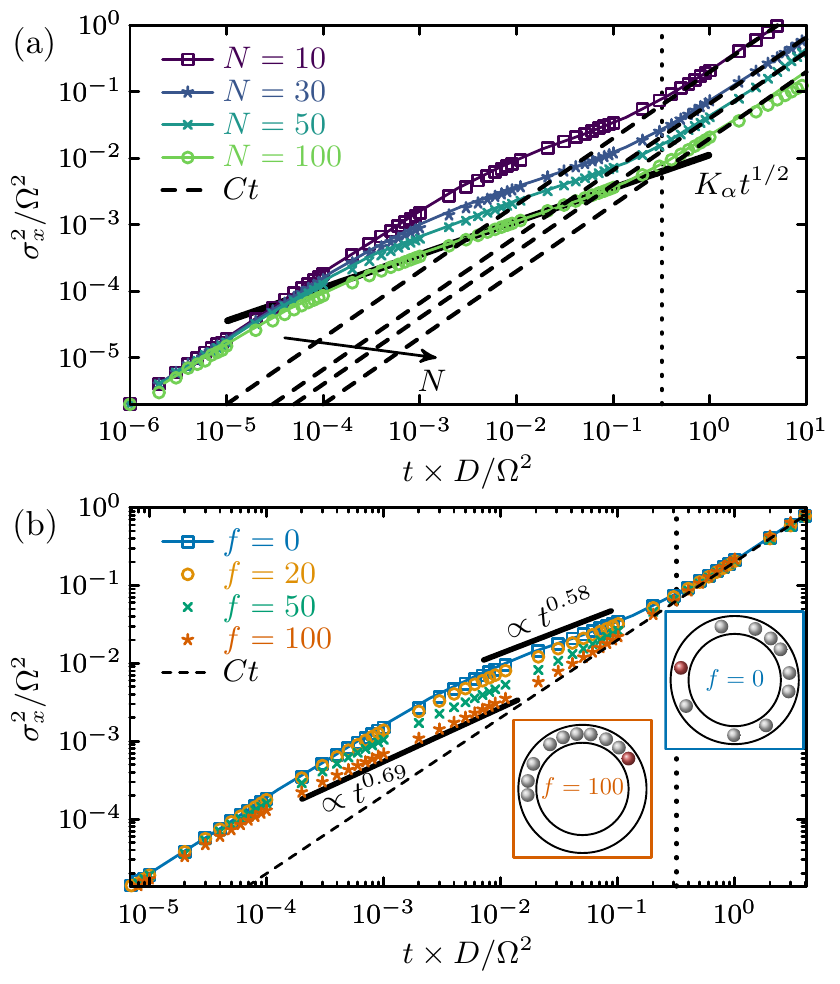}
 \caption{Variance of particle-displacement in a
   single file on a ring (see Fig.~\ref{fig:anomalous_finite}a). (a)~All $N$ particles are pulled
   by a Force $F$ (here $F=0$); (b) only the tagged particle
   is pulled by a force $F\Omega\equiv f\times k_{\rm B}T$ (the inset
   depicts the effect of $F$) with $N=10$.
 Symbols represent the centralized TAMSD \cite{Note1} extracted
 from a long trajectory $\tau=10^3\times D/\Omega^2$ for each $N$.  The lines are deduced from a modified Jepsen mapping (see SM \cite{Note3}).
 Parameters: $D=k_{\rm B}T=\Omega=1$ and $d=0$, i.e.,  time is measured in units of $D/\Omega^2$ and displacements in units of $\Omega=l-Nd$.
  }
 \label{fig:SF}
\end{figure}

\emph{Driven single file on a ring.---}We first consider a single file of $N$ impenetrable Brownian
particles with diameter $d$
and a diffusion coefficient $D$
\emph{all} dragged with a constant force $F$
described by 
the Langevin equation $\dot x_i(t)=\gamma^{-1}F+\xi_i(t)$ for $i=1,\ldots N$,
where the friction coefficient obeys
the fluctuation-dissipation relation $\gamma=k_{\rm B}T/D$
and $\xi_i(t)$ 
represents Gaussian white noise with zero mean and
covariance 
$\avg{\xi_i(t)\xi_j(t')}=2D\delta_{ij}\delta (t-t')$.
The
hardcore interaction imposes internal boundary conditions
$x_i<x_{i+1}+d$ and the confinement to a ring with circumference $l$
(see Fig.~\ref{fig:anomalous_finite}a) additionally imposes 
$x_N-x_1\le l-d$, i.e., the first particle blocks the passage of
the last one.
We refer to this setting as ``pseudo non-equilibrium'' since the
transformation to a coordinate system rotating with velocity
$v=\gamma^{-1}F$ virtually restores
equilibrium dynamics with vanishing current \cite{Note3}. 
Nevertheless, the power required to drag the $N$ particles with velocity
$v=\gamma^{-1}F$ against the friction force is $\dot W^{\rm
  ss}=N\times Fv$ and Eq.~\eqref{eq:TURdef} in turn yields
$C=2k_{\rm B}T/\gamma N$, a result independent of $F$ (see also \cite{nels99}).

It is well known that a tracer particle in a dense single-file ($1\ll N<\infty$)
exhibits transient subdiffusion according to Eq.~\eqref{eq:anomalous} with
exponent $\alpha\simeq1/2$ and generalized diffusion 
constant $K_\alpha\simeq2N^{-1}\Omega\sqrt{D/\pi}$ (see, e.g. \cite{harr65,lin05,talo06,liza08,liza09,liza10,delf11,leib13,krap14,ryab16}
and experiments in \cite{hahn96,wei00,lutz04}), where
$\Omega\equiv l-Nd$ is the free volume on a ring with circumference $l$.
Therefore, the inequality \eqref{eq:TURdef} implies that subdiffusion
can persist at most until a time $t^*=(K_\alpha/C)^{1/(1-\alpha)}\simeq\Omega^{2}/(D\pi)$ (see vertical line in Fig.~\ref{fig:SF}a and Eq.~\eqref{eq:tanomalous}).

Thermodynamic consistency limits the extent of subdiffusion to time-scales
$t\lesssim t^*$. To test the bound in Fig.~\ref{fig:SF}a we determined the centralized TAMSD (see symbols)
of a tracer particle
from a single trajectory of length $\tau=10^3\times (D/\Omega)$ generated by a Brownian dynamics simulation 
with time increment $d t=10^{-6}\times (D/\Omega)$, and independently
deduced $\sigma_x^2$ also from a mapping inspired by Jepsen
\cite{jeps65} (see lines, SM \cite{Note3} as well as \cite{evan79,cool05}). 
The results confirm that the TUR sharply
bounds the duration of subdiffusion terminating at time $t^*$ (see
intersection of the TUR-bound and vertical line). 
If we were to allow particles to overtake the long-time asymptotics would not saturate at the dashed line
(see Fig. 10(a-c) in \cite{luce12}) --- in this scenario subdiffusion may terminate before $t^*$.

\emph{Active single file.---}A ``genuinely'' non-equilibrium steady state is generated by
pulling only the tagged particle with a force $F$. The tagged-particle
diffusion quantified by $\sigma_x^2(t)$ is shown in Fig.~\ref{fig:SF}b.
Here the non-equilibrium driving force $f\equiv F\times \Omega/k_{\rm B}T$
increases the anomalous exponent from $\alpha\approx 0.58$
to $\alpha\approx 0.69$.
Nevertheless, the TUR (dashed line) still tightly bounds the time subdiffusion
terminates. Moreover, the onset of subdiffusion is shifted towards 
shorter times which
may be explained as follows.
A strongly driven particle ``pushes'' the non-active particles thereby locally
increasing density
which in turn shifts the onset of subdiffusion.
The effect increases with the
strenght of the driving (see inset ``$f=100$'' in Fig.~\ref{fig:SF}b).
This result seemingly contradicts
previous findings on active lattice models at \emph{high density} showing that all even cumulants (incl. the variance)
remain unaffected by the driving $f$ \cite{illi13} (see also \cite{beni13}).
The contradiction is only apparent --- single file diffusion for any number of particles in fact corresponds
to the \emph{low density} limit of lattice exclusion models.

\begin{figure}
  \includegraphics{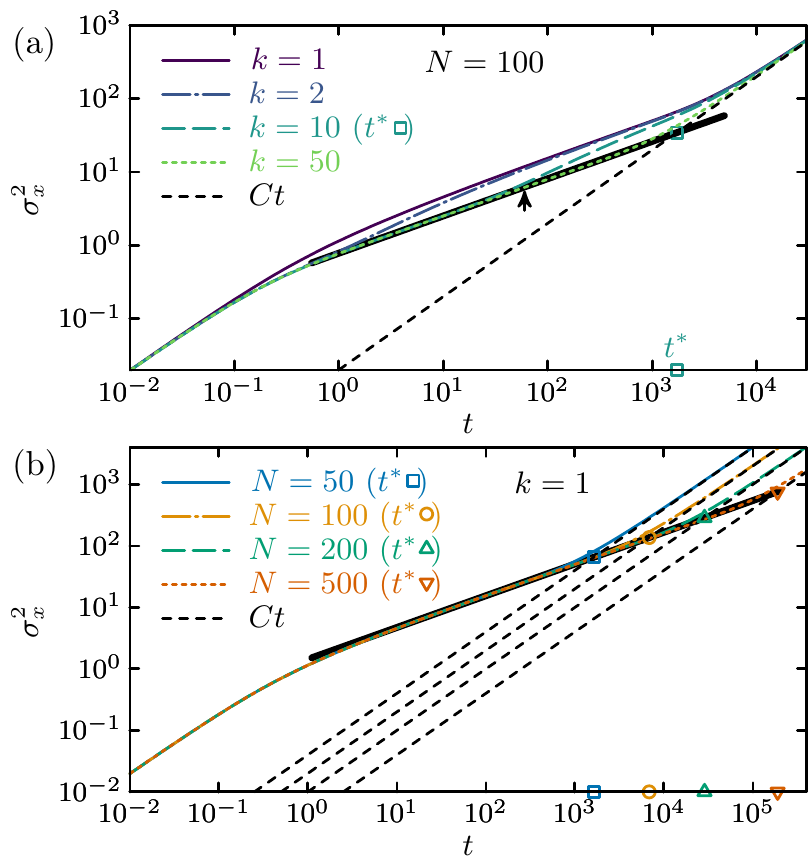}
 \caption{(a) $\sigma_x^2(t)$ from Eq.~\eqref{eq:rouse_exact} for a dragged
   Gaussian chain with
   $N=100$ beads, where we tag the $k$th particle ($k=1,2,10,50$). The
   TUR-bound is shown as the dashed black line. Taking e.g. $k=10$ 
 we find transient subdiffusion $\sigma_x^2(t)\simeq K_\alpha t^\alpha$
 (solid black line) in the vicinity of $t\sim t_{\rm ref}\equiv 10^1$;
 using Eq.~\eqref{eq:rouse_exact} yields the exponent $\alpha\equiv t\del_t\ln\sigma^2(t)|_{t=t_{\rm ref}}\approx
0.508$ with $K_\alpha\equiv t_{\rm
  ref}^{-\alpha}\sigma^2(t_{\rm ref})$. The rectangle denotes the upper bound
on the extent of subdiffusion $t^*$ while the vertical arrow
highlights the actual time at which the subdiffusive regime for $k=10$
terminates. (b) $\sigma_x^2(t)$ of the first bead ($k=1$) for
 increasing $N$. Symbols denote the TUR-bound.}
 \label{fig:rouse}
\end{figure}

\emph{Gaussian chain (Rouse model).---}We now consider a harmonic
chain with $N$ beads
(see 
Fig.~\ref{fig:anomalous_finite}b). The equations of motion (for the
time being in absence of a pulling force) correspond to
\cite{rous53,fugm10,wutt11}
$\dot x_k(t)=-D\sum_lH_{kl}x_l(t)+\xi_k(t)$ where
$(\mathbf{H})_{kl}=H_{kl}$ is the
Hessian of 
$U=\sum_{i=2}^N(x_{i}-x_{i-1})^2/2$.
We set $\gamma^{-1}=D$, i.e., $k_{\rm B}T\equiv 1$. 
The variance of the  $k$th bead's position reads (see e.g. \cite{gard04}) 
\begin{equation}
\sigma_{x}^2(t)=\frac{2}{N}\left[Dt\!+\!\sum_{p=1}^{N-1}\cos^2\!\left(\frac{\pi p(2k-1)}{2N}\right)\frac{1-\e^{-2D\lambda_p t}}{\lambda_p}\right]\!,
\label{eq:rouse_exact}
\end{equation}
where  $\lambda_p=4\sin^2(\pi p/2N)$ \cite{rous53,fugm10,wutt11}.
The first term in Eq.~\eqref{eq:rouse_exact} corresponds to the
center-of-mass diffusion. 

Suppose now that we drag \emph{all} particles with a constant force $F$. In this case the force affects only
the mean displacements but \emph{not} the variance \cite{Note3}. 
In other
words,  the left hand side of Eq.~\eqref{eq:TURdef}
is not affected by 
$F$, whereas the right hand side
becomes $C=2D/N$ since $\dot{W}_{\rm ss} = v\times NF$ with
$v=\gamma^{-1}F=DF$. By inspecting Eq.~\eqref{eq:rouse_exact}
directly (note that all terms in
Eq.~\eqref{eq:rouse_exact} are non-negative) one can verify that the TUR indeed bounds
the diffusion of the $k$th particle by
$\sigma_{x}^2(t)\ge 2D t/N$ at any time $t$.   

In Fig.~\ref{fig:rouse}a we inspect the sharpness of the bound. For
example, tagging the $10$th bead in 
a polymer with $N=100$ we observe subdiffusion with an
exponent $\alpha\approx 0.508$ (see thick black line) that terminates at
$t<t^*$ (see vertical arrow), i.e. faster than predicted by the TUR
(see green rectangle). Interestingly, the scaling of $\sigma_{x}^2(t)$
at this point does not become normal with $\alpha=1$ but instead turns to a
second, slightly larger anomalous exponent. Normal diffusion is in
fact observed at much longer times.
This example highlights that subdiffusion with an (initial) exponent
$\alpha$ cannot extend beyond $t^*$. However, this does \emph{not}
imply that $t^*$ necessarily corresponds to the onset of normal diffusion. 
Conversely, if we tag the first particle of the chain (see Fig.~\ref{fig:rouse}b)
the TUR bounds the overall duration of subdiffusion
quite tightly. According to Eq.~\eqref{eq:tanomalous} the longest time
subdiffusion can persist  increases with $N$ as $t^*\propto
C^{-2}\propto N^2$ (see symbols in
Fig.~\ref{fig:rouse}b as well as \cite{hoef13}).

\emph{Superdiffusion in the active comb model.---}So far we have
discussed only systems exhibiting subdiffusion. To address
superdiffusion we consider the ``active comb model'' depicted in
Fig.~\ref{fig:anomalous_finite}c corresponding to diffusion on a ring
with side-branches with a \emph{finite} length $L$ oriented perpendicularly to the ring at positions separated by $l$. Within the
ring (but not in the side-branches) 
the particle is dragged with a constant
force $F$.
For simplicity we assume the diffusion constant, $D$, to be the same
in the ring and along the side-branches. 
The probability density and flux are assumed to be continuous at the
intersecting nodes such that the steady state probability to find the
particle in the ring (i.e. in a ``mobile state'') corresponds to
$\phi_m=l/(l+2L)$ yielding a mean drift velocity $v=\beta D F
\phi_m$. Using $\dot W_{\rm ss}=Fv$ alongside the TUR
(Eq.~\eqref{eq:TURdef}) we immediately obtain  $\sigma_x^2(t)\ge 2 \phi_m Dt$.
It is known that infinite side-branches ``$L=\infty$'' in the passive
comb model (i.e. $F=0$) break ergodicity. That is, a
non-equilibrium steady state ceases to exist and subdiffusion with
exponent $\alpha=1/2$  persists for any fixed initial condition and
time $t$ (e.g., see \cite{bouc90,bere14,beni15}). 
Conversely, a bias  $F\neq0$ in a finite comb ($L<\infty$) was
found, quite counterintuitively, to enhance the long time diffusion \cite{bere15}, which leads to transient superdiffusion as discussed below.

The particle's position along
the ring does not change while it 
is in a side-branch.  Therefore, 
only
the (random) ``occupation time in the mobile phase'' \cite{rebe13,lapo20},
$\tau_m(t)\le t$, 
is relevant.
Its fraction is 
referred to as
the ``empirical density'' \cite{bara15c,lapo20}
since 
$\avg{\tau_m(t)}=\phi_mt$.

\begin{figure}
\includegraphics{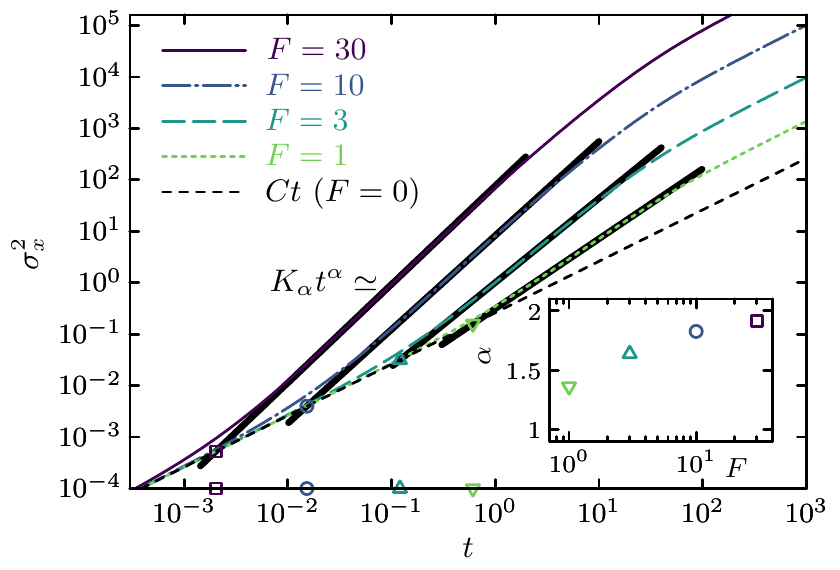}
 \caption{$\sigma_x^2$ in the driven comb model (see Fig.~\ref{fig:anomalous_finite}d). We consider various driving
 forces $F$ and side-branches with
   length $L=10$ separated by a distance $l=3$ yielding a steady state
   probability in the ring $\phi_m=l/(l+2L)=3/23\approx0.13$ with $D=\beta=1$.
 The force-free case $F=0$ coincides with the bound $Ct$ in
 Eq.~\eqref{eq:TURdef}.
 The thick
 lines correspond to $K_\alpha t^\alpha$ with the ``maximal exponent'' $\alpha\equiv\max_t t\del_t\ln \sigma_x^2(t)$ depicted in the inset.
Symbols denote the time $t^*$ in Eq.~\eqref{eq:tanomalous} where
$K_\alpha t^\alpha$ and $Ct$ intersect. Long times $t\to \infty$
and strong driving $\beta Fl\gg 1$ yield
$\sigma_{x}^2\simeq 2D\phi_mt+(\beta F l)^2(1-\phi_m)^3Dt/6$.
 }
 \label{fig:comb}
\end{figure}

The particle drifts
with velocity  $\beta D F$  and diffuses with diffusion constant
$D$ during the time $\tau_m(t)$ it spends in the ring. This implies a
displacement 
distributed according to $x_t\sim \beta D F\tau_m(t)+\sqrt{2
  D\tau_m(t)}\mathcal{N}$, where $\mathcal{N}$
is a standard normal random number, which eventually leads to (for an
alternative derivation see \cite{bere15})
\begin{equation}
 \sigma_{x}^2(t)
 =2 D\phi_m t+(\beta D F)^2\sigma_{\tau}^2(t),
 \label{eq:comb_exact}
\end{equation}
where we used $\avg{\mathcal{N}^2}=1$, $\avg{\mathcal{N}}=0$, $\avg{\tau_m(t)}=\phi_m t$ 
and defined $\sigma_{\tau}^2(t)\equiv\avg{\tau_m(t)^2}-\avg{\tau_m(t)}^2$.
To deduce $\sigma_{\tau}^2(t)$ we translated the equation of motion
into a Markov jump system according to \cite{holu19}
and used a spectral expansion \cite{lapo20} which alongside
Eq.~\eqref{eq:comb_exact} yields $\sigma_x^2(t)$. The result for
$l=3$  and $L=10$ \cite{Note3} is shown in Fig.~\ref{fig:comb}. The
thick lines denote power laws with a ``maximal exponent''
$\alpha=\max_t t\del_t \ln \sigma_x^2(t)$ (see inset for the respective
values). At equilibrium ($F=0$) the diffusion is normal at all
times. The presence of a force causes transient
superdiffusion with an exponent approaching the ballistic regime
$\alpha\approx2$ upon increasing $F$. Note that here the TUR bounds
the \emph{time of initiation} of superdiffusion (see symbols) and \emph{not} the termination.

To explain this we must understand when
$\sigma_{\tau}^2(t)$ increases non-linearly with $t$.
One can show that for sufficiently small times $t\to 0$ the particle is found
with high probability 
either only in the ring or only in one of the side-branches
which yields a vanishing variance
$\sigma_{\tau}^2(t)=\mathcal{O}(t)$. Conversely, we have recently
found \cite{lapo20} that the dispersion of the fraction of occupation
time at long times,
$\mathcal{D}\equiv \lim_{t\to \infty}\sigma_{\tau}^2(t)/t$, is
entirely encoded in the (steady state) joint return probability, $P(m,t,m)$, i.e.
the probability to be in the mobile region $m$ initially and again at time $t$
\begin{align}
\mathcal{D}&=2\int_0^\infty [P(m,t,m)-\phi_m^2]\dd t\nonumber\\
&=
\frac{4lL^2[(\beta F)^2lL+3\beta Fl\coth(\beta Fl/2)-6]}{3D(\beta F)^2(l+2L)^3},
\label{eq:Dinf}
\end{align}
where the first line is shown in \cite{lapo20},
and the second line is derived in \cite{Note3} (a
similar result is found in \cite{bere15}).
At strong driving $\beta F l \gg 1$ we find
$\mathcal{D}\simeq l^2(1-\phi_m)^3/6 
$
which interestingly enhances diffusion $\propto (\beta
F l)^2(1-\phi_m)^3$ by a magnitude that increases with the likelihood to
reside immobile.
Superdiffusion thus arises from an interplay between
effectively ``ballistic'' transport in the ring and pausing
in the side-branches, and  
becomes pronounced at strong driving $\beta F l\gg 1$
and in the presence of long side-branches $L\gg l$, yielding $1-\phi_m\approx 1$. 
 A similar effect gives rise to the so-called Taylor dispersion
 \cite{tayl53} that occurs in diffusion in a flow field \cite{broe87,kahl17,aure17}.

\emph{Conclusion.---}We established a bridge between anomalous
diffusion and the TUR by explaining how the latter can be utilized to (sharply)
bound the temporal extent of anomalous diffusion in finite systems
driven out of equilibrium. We used the TUR to demonstrate that a
non-equilibrium driving may in fact be required for anomalous
dynamics to occur such as e.g. in the comb model. 
We have shown that the
TUR can also bound the duration of
anomalous diffusion in systems obeying detailed balance if we
are able to construct a fictitious non-equilibrium system with the
same dynamics, which we demonstrated by means of the passive and
driven single file and the Rouse polymer. In this context it will be
useful to deepen the connection between the TUR \cite{maci18} and
anomalous transport \cite{lutz01,gode13} close to equilibrium, growing interfaces \cite{nigg20,nigg21},  and to
bound subdiffusion in flexible gel networks \cite{gode14}.

Finally, we point out that the TUR (Eq.~\eqref{eq:TURdef}) and therefore our results apply
to \emph{overdamped systems}  (i.e., when momenta relax
``instantaneously''). If we include momenta or
consider the presence of magnetic
fields the TUR requires modifications \cite{proe19,chun19}. Such
extensions will allow to
bound the extent of anomalous
diffusion in underdamped systems \cite{metz02,buro08,buro08a,goyc19,goyc20}. Finally, the recent generalization of the TUR \cite{liu20,koyu20,dech18} will allow applying the TUR to anomalous diffusion and anomalous displacements arising from non-stationary and non-ergodic infinite 
 systems \cite{illi13}.

\begin{acknowledgments}
 The financial support from the German Research Foundation (DFG) through the Emmy Noether Program GO 2762/1-1 to A. G. is gratefully acknowledged
\end{acknowledgments}

\renewcommand{\theequation}{S\arabic{equation}}
\renewcommand{\thefigure}{S\arabic{figure}}

\appendix

\section*{Supplemental Material}
In this Supplemental material we first clarify why the long time diffusion does not suffice to bound the extent of anomalous subdiffusion (see Sec.~\ref{sec:Dinf_vs_TUR}). 
In Sec.~\ref{sec:pseudo} we show that the thermodynamic uncertainty
relation can also be applied to
interacting many particle systems at equilibrium if one can construct/identify a ``pseudo non-equilibrium state''. We explain the mapping motivated by Jepsen that we used to analyze the single-file system (lines in Fig. 2 in the main text) in Sec.~\ref{sec:SF_appendix}. In Sec.\ref{sec:comb} we derive the second line in
Eq. (6) in the main text and provide details about the numerical implementation of the comb-model (see Fig. 4 in the main text).

\subsection{Long time diffusion does not bound the extent of anomalous
diffusion}\label{sec:Dinf_vs_TUR}
\begin{figure}
 \includegraphics{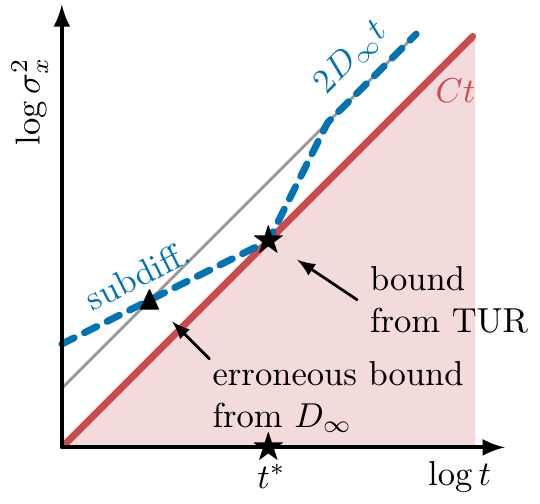}
 \caption{Knowing the long time diffusion $D_\infty$ alone does not suffice to bound the extent of subdiffusion.
 }
 \label{fig:SI_whyTUR}
 \end{figure}

In the limit of ergodically long times the variance will grow linearly
in time with a diffusion coefficient $D_\infty$ which follows from
$\sigma_x^2(t)/(2t)\to D_\infty$ or $\sigma_x^2(t)\simeq 2 D_\infty t$
in the limit $t\to \infty$. Note that the limit $\sigma_x^2(t)/(2t)\to D_\infty$ does not exclude the possibility
of approaching $D_\infty t$  from below. Therefore, knowing
$D_\infty$ alone cannot suffice to bound the extent of subdiffusion.
To illustrate this we consider an approach of the long time diffusion from below as shown in Fig.~\ref{fig:SI_whyTUR}.
From the long time asymptotics we would ``erroneously'' underestimate the latest end of subdiffusion with a constant exponent $\alpha <1$  (see triangle).
The inequality ``$\ge$'' involving the thermodynamic uncertainty
relation (TUR), $\sigma_x^2(t) \ge Ct$, prevents any approach to normal
diffusion to intermediately cross the red line. Thus $t^*$ is always guaranteed to be the latest
time when subdiffusion with a constant anomalous exponent must end.

 \subsection{Uncertainty relation at equilibrium -- pseudo non-equilibrium}\label{sec:pseudo}
In this section we explain how the thermodynamic uncertainty
relation obeyed by non-equilibrium systems,
\emph{under given conditions} can also be applied to interacting colloidal particles at equilibrium. 
To this end we consider translationally invariant systems that are at
equilibrium with zero drift velocity $v=0$. Adding a drift $\alpha
=vt$ to \textbf{\emph{all}} coordinates transforms to system to what we call ``pseudo non-equilibrium''. In the following two paragraphs we explain that the \emph{drifting pseudo non-equilibrium} and the corresponding \emph{non-drifting equilibrium}
system display the same variance
$ \sigma_x^2(t)\big|_{v=0}=\sigma_x^2(t)\big|_{v\neq0}$.

Let us first explain this idea mathematically. Consider a random
variable $X$ and a shifted random variable $Y+\alpha$, where $\alpha$
is some constant. The variance is known to be 
invariant
with respect
to such a constant bias, i.e.\ $\operatorname{Var}(Y)=\operatorname{Var}(X+\alpha)=\operatorname{Var}(X)$. In the following
we prove this mathematical property in the context of physically
interacting particles. 

Consider $N$ colloidal particles interacting via a pairwise additive interaction potential $U(\boldsymbol{x})=\sum_{i<j} U_{ij}(x_i-x_j)$, where $\boldsymbol{x}=(x_1,\ldots,x_N)=\boldsymbol{x}(t)$ is a vector with all particle positions at time $t$. 
At equilibrium the 
colloidal particles obey the set of coupled Langevin equations
\begin{equation}
 \dot x_i(t)=-\gamma^{-1}\partial_{x_i}U(\boldsymbol{x})+\xi_i(t)
 \label{eq:F=0},
\end{equation}
where
$\xi_t$ is Gaussian white noise with zero mean and covariance $\langle \xi_i(t)\xi_j(t')\rangle=2\delta_{ij}\delta(t-t')$.
We now add a drift to \emph{all particle positions} such that $y_i\equiv x_i(t)+vt$ (for all $i$).
The constant velocity may arise from a drifting coordinate system with
constant velocity $v$. In this case the system is only fictitiously
driven out of equilibrium \cite{spec08} to which we refer as ``pseudo non-equilibrium''.
Note that  the same ``pseudo non-equilibrium'' state
may also be generated by
a ``real'' physical force $F$, i.e.\ $v=\gamma^{-1}F$.
In both cases the drifted colloidal particles satisfy the Langevin equations
\begin{equation}
 \dot y_i(t)=v-\gamma^{-1}\partial_{x_i}U(\boldsymbol{x})+\xi_t=\gamma^{-1}F-\gamma^{-1}\partial_{y_i}U(\boldsymbol{y})+\xi_t,
  \label{eq:f!=0}
\end{equation}
where in the last step we defined $\gamma^{-1}F\equiv v$ and used $U_{ij}(y_i-y_j)=U_{ij}(x_i+vt-x_j-vt)=U_{ij}(x_i-x_j)$.
Eq.~\eqref{eq:f!=0} establishes that  $y_i\equiv x_i(t)+\gamma^{-1}Ft$ connects equilibrium dynamics
$\{x_i(t)\}$ to a fictitiously dragged system $\{y_i(t)\}$.
For any fixed $\tau$ the displacement after time $t$ using $\delta x_i(t)\equiv x_i(t+\tau)-x_i(\tau)$,  becomes
$\delta y_i(t)\equiv y_i(t+\tau)-y_i(\tau)=\delta x_i(t)+\gamma^{-1}F t$, which 
yields the variance
\begin{align}
 \sigma_y^2(t)&\equiv\langle\delta y_i(t)^2\rangle-\langle\delta y_i(t)\rangle^2\nonumber\\
& =\langle[\delta x_i(t)+\gamma^{-1}Ft]^2\rangle-[\langle\delta x_i(t)\rangle+\gamma^{-1}Ft]^2\nonumber\\
 &=\langle\delta x_i(t)^2\rangle-\langle\delta x_i(t)\rangle^2=\sigma_x^2(t).
 \label{eq:varConst}
\end{align}
Thus the driven ($F\neq0$) and equilibrium ($F=0$) system have exactly
the same variance.

In the pseudo non-equilibrium state \eqref{eq:f!=0} the dissipation rate becomes $\dot{W}_{\rm ss}=NFv=N\gamma v^2$. Using the fluctuation dissipation relation $\gamma=k_{\rm B}T/D$ yields 
 $\dot{W}_{\rm ss}=Nk_{\rm B}T/D$ which with Eq.~(2) in the main text yields $C=2D/N$ -- a result that is independent of the force. This result in conjunction with the preservation 
 of variance, Eq.~\eqref{eq:varConst}, in turn yields
\begin{equation}
 \sigma_x^2(t)\big|_{F=0}= \underbrace{\sigma_y^2(t)\big|_{F\neq0}\ge  C t\big|_{F\neq0}}_{\text{TUR}}=C t\big|_{F=0},
 \label{eq:TUReq}
\end{equation}
which completes the proof that the TUR may be applied to equilibrium
systems for which we are able to construct a pseudo non-equilibrium
that displays the same variance.

We employed the TUR according to Eq.~\eqref{eq:TUReq} for the derivation of the results depicted in Fig.~2a (single-file) and Fig.~3 (Gaussian chain). The comb model from Fig. 1c in the main text is not translation invariant due to the side-branches with length $L$, which is why the the result depicted in Fig. 4 in the main text cannot be studied in this manner. Moreover, the driving force in the active single file (Fig.~2b in the main text) affects the interaction between particles (see insets). Thus Eq.~\eqref{eq:TUReq} is bound to hold only if both the system is translational invariant \emph{and} the same biasing force is applied to \emph{all} particles. In the following subsection we discuss a translational invariant system -- the single file.

\subsection{Single file}\label{sec:SF_appendix}

\emph{Jepsen mapping in the single file.} 
For simplicity we set the length of the ring to $l=1$ and diameter $d=0$ such that the particle position $\boldsymbol{\theta}=(\theta_1,\ldots,\theta_N)$ will satisfy 
\begin{equation}
 \theta_1(t)\le\theta_2(t)\le\ldots\le \theta_N(t)\quad\text{and}\quad\theta_N(t)-\theta_1(t)\le 1.
\label{eq:condition_single_file}
 \end{equation}
Without loss of generality we keep $d=0$ and $l=1$.
Note that the problem of having particles with finite diameter $d>0$ moving on ring with circumference $l$ can be restored easily via the mapping $\theta_i(t)\to (l-Nd)\theta_i(t)+(i-1)d$.

The first expression in Eq.~\eqref{eq:condition_single_file} means that the particles cannot penetrate each other such that the order is preserved, and the last condition is due to the ring-like structure and means that the last particle  -- the $N$th one --
cannot advance the first particle by more than one circumference $l=1$.
Let us now consider the method developed by Jepsen \cite{jeps65} (e.g., see also Refs.~\cite{evan79,cool05,bark10}), 
which allows us to map the system of interacting particles through
Eq.~\eqref{eq:condition_single_file} onto a system of non-interacting
particles $\tilde{\boldsymbol{\theta}}$ that may violate Eq.~\eqref{eq:condition_single_file}.
Jepsen \cite{jeps65} derived a mapping which restores the first expression in Eq.~\eqref{eq:condition_single_file} by permuting the particles positions ``$\operatorname{sort}(\cdots)$'' into increasing order such that $\operatorname{sort}(\tilde{\boldsymbol{\theta}})$
satisfies the first condition in \eqref{eq:condition_single_file}, i.e.,
\begin{equation}
 \operatorname{sort}[\tilde{\boldsymbol{\theta}}]_1\le \operatorname{sort}[\tilde{\boldsymbol{\theta}}]_2\le\ldots\le \operatorname{sort}[\tilde{\boldsymbol{\theta}}]_N.
 \label{eq:sort}
\end{equation}
To also restore the second condition we need to go beyond Ref.~\cite{jeps65}
and find a map
$\mathcal{M}(\tilde{\boldsymbol{\theta}})= \boldsymbol{\theta}$,
which also restores the periodic boundary condition in
\eqref{eq:condition_single_file}. That is, we need to find the mapping $\mathcal{M}$ that fully restores Eq.~\eqref{eq:condition_single_file}.
Introducing the element-wise floor function $\lfloor\cdot\rfloor$, adopting the sorting function \eqref{eq:sort},
defining the
mean value of a vector $\by$ through $\overline{\by}\equiv N^{-1}\sum_{i=1}^Ny_i=\overline{\operatorname{sort}(\by)}$,
the
desired mapping $\mathcal{M}$ is given by
 \begin{equation}
 \begin{pmatrix}
 \theta_1\\
 \vdots\\
 \theta_{N-\nu}\\
 \theta_{N-\nu+1}\\
 \vdots\\
 \theta_N
 \end{pmatrix}
 =\mathcal{M}(\tilde{\boldsymbol{\theta}})=
 \begin{pmatrix}
  \operatorname{sort}(\tilde{\boldsymbol{\theta}}-\lfloor\tilde{\boldsymbol{\theta}}\rfloor)_{\nu+1}+\vartheta\\
  \vdots\\
  \operatorname{sort}(\tilde{\boldsymbol{\theta}}-\lfloor\tilde{\boldsymbol{\theta}}\rfloor)_{N}+\vartheta\\
  \operatorname{sort}(\tilde{\boldsymbol{\theta}}-\lfloor\tilde{\boldsymbol{\theta}}\rfloor)_{1}+\vartheta+1\\
  \vdots\\
  \operatorname{sort}(\tilde{\boldsymbol{\theta}}-\lfloor\tilde{\boldsymbol{\theta}}\rfloor)_{\nu}+\vartheta+1
 \end{pmatrix},
\label{eq:Jepsen_mapping}
\end{equation}
where
\begin{equation}
 \vartheta\equiv\lfloor\overline{\lfloor\tilde{\boldsymbol{\theta}}\rfloor}\rfloor
 \quad\text{and}\quad
\nu=N \overline{\lfloor\tilde{\boldsymbol{\theta}}\rfloor}-N \vartheta.
\label{eq:Jepsen_mapping2}
\end{equation}
The mapping \eqref{eq:Jepsen_mapping} can be shown to restore
Eq.~\eqref{eq:condition_single_file} entirely
and to preserve the total displacement $\sum_{i=1}^N\theta_i(t)=\sum_{i=1}^N\tilde\theta_i(t)$.
Note that the variable $\vartheta$ (and $\vartheta+1$) in Eq.~\eqref{eq:Jepsen_mapping} counts the number complete revolutions in the ring $\vartheta+\nu$.

Without loss of generality we tag particle 1 such that
the displacement of the tagged particle becomes $x_t=\theta_1(t)-\theta_1(0)$. Note that a particle with non-zero diameter $d$ and ring with length $l$ (incl. $l\neq1$) can be accounted for  by  $x_t=\Omega[\theta_1(t)-\theta_1(0)]$, where $\Omega=l-Nd$.

\textit{Numerical evaluation of the variance $\sigma_x^2(t)$.} We determine $\sigma_x^2(t)$ at any distant time $t$ by directly evaluating $10^4$ realizations of positions $x_t$ at time $t$. Each position $x_t$
is generated as follows (we set $d=0$, $\Omega=l=1$).
\begin{itemize}
\item[(i)] Distribute all particles $i=1,\ldots,N$ uniformly on the scaled ring via $u_i\sim \operatorname{uniform}[0,1]$ and sort them $\tilde\theta_i(t)=\operatorname{sort}(\boldsymbol{u})_i$.

 \item[(ii)]  Generate $N$ standard normal Gaussian random variables $ Z_1,\ldots, Z_N$ and propagate the scaled coordinate $\tilde\theta_i(t)=\tilde\theta_i(0)+\Omega^{-1}[\gamma^{-1} F t+\sqrt{2 D t}\cdot Z_i]$ for $i=1,\ldots,N$
 \item[(iii)] Restore the order of particles by backtracking \emph{all} collisions, which is attained by  evaluating $\boldsymbol{\theta} (t)=\mathcal{M}(\tilde{\boldsymbol{\theta}}(t))$ from Eqs.~\eqref{eq:Jepsen_mapping} and \eqref{eq:Jepsen_mapping2}. Note that $\boldsymbol{\theta} (0)=\mathcal{M}(\tilde{\boldsymbol{\theta}}(0))=\tilde{\boldsymbol{\theta}}(0)$.
 \item [(iv)] One realization of the displacement of the tagged particle is obtained from
 $x_t=\Omega[\theta_1 (t)-\theta_1 (0)]$. Here $\Omega=1$.
\end{itemize}

A realization of the displacement $x_t$ after any time $t$ is obtained
according to steps (i)-(iv). For each time $t$ in Fig.~2a in the Letter we evaluated (see lines)
$10^4$ displacements $x_t$ and deduced their corresponding variance. 
In contrast to the Brownian Dynamics simulation (see symbols) the steps (i)-(iv) avoid any intermediate time step in the simulation.
Note that this efficient mapping can only be used when all particles
are dragged by the same force (see Fig.~2a in the main text). As soon
as only one particle is dragged as depicted in Fig.~2b in the main
text the method developed by Jepsen \emph{cannot} be employed.

\subsection{Comb model}\label{sec:comb}

\subsubsection{Comb model dynamics}
As explained in the main text it suffices to merely focus on the stochastic time $\tau_m(t)$ spend in the ``mobile'' ring region.
Since the comb is assumed to be periodic we merely focus on one period from one pair of side-branches to the next.
We call $P_0(x,t)$ the probability density to find the particle in the
``mobile'' ring at distance $x$ along the force $F$ from the
``previous'' side branch. We further denote  the
probability density within any of the two side-branches
to be at time $t$ the distance $y$ away from the ``mobile'' ring by $P_y(0,t)$.
The probability density $P_y(x,t)$ satisfies $\int_0^L P_y(0,t)\dd y+\int_0^l P_0(x,t)\dd y=1$.
The Fokker-Planck equation then reads
\begin{equation}
\begin{aligned}
 \del_t P_y(0,t)&=D \del_y^2P_y(0,t),\\
 \del_t P_0(x,t)&=-D[\beta F\del_x-\del_x^2 ]P_0(x,t),
\end{aligned}
\label{eq:FPE_comb_time}
\end{equation}
where $0\le x\le l$ and  $0\le y\le L$, along 
with the boundary condition $P_0(0_+,t)=P_0(l-0_+,t)=P_{0_+}(0,t)/2$ (division by 2 accounts for two side-branches) and conservation of probability $P_0(0_+,t)+\del_xP_0(0_+,t)=\del_xP_0(l-0_+,t)$ as well as $\del_yP_L(0,t)=0$.
Eq.~\eqref{eq:FPE_comb_time} corresponds to a diffusion on a piece-wise one dimensional graph \cite{hart20arxiv}.
The stationary state probability density  becomes $P_0(x,\infty)=1/(2L+l)$ and $P_y(0,\infty)=2/(2L+l)$.
The mean occupation time in the mobile state becomes $\avg{\tau_m(t)}=t\int\dd xP_0(x,\infty)=lt/(2L+l)=\phi_m t$, where $\phi_m\equiv l/(2L+l)$. In the remainder of this section we determine the variance of the occupation time $\sigma_\tau^2(t)=\avg{\tau_m(t)^2}-\avg{\tau_m(t)}^2$.

\subsubsection{Numerical solution of the driven comb model}
To solve the Fokker-Planck equation numerically according to Ref.~\cite{holu19} we translate the partial differential equation into a master equation, i.e., a random walk on a discrete grid with equidistant spacing $\delta$ such that the ring states
 $z=0,1, \ldots, N_1$ ($N_1=l/\delta$) correspond to positions
$x=0,\delta, \ldots, \delta N_1$, while states belonging to the side-branches separated by $\delta$
are $z=0,(N_1+1),(N_1+2) , \ldots,(N_1+N_2) $ and correspond to the positions 
$y=0,\delta , \ldots,N_2\delta $ ($N_2=L/\delta$). For convenience we assume that both $l/\delta$ and $L/\delta$ are integer valued.
The transition rates within the ``ring states'' $z\in\{0,1,\ldots N_1\}$
are given by
\begin{equation}
\begin{aligned}
 R_{z\to z+1}&=\frac{D\e^{\beta DF\delta /2}}{\delta^2},\quad\text{for $0\le z\le N_1-1$}\\
  R_{N_1\to 0}&=\frac{D\e^{\beta DF\delta /2}}{\delta^2},\quad  R_{0\to N_1}=\frac{D\e^{-\beta DF\delta /2}}{\delta^2},\\
  R_{z\to z-1}&=\frac{D\e^{-\beta DF\delta /2}}{\delta^2},\quad\text{for $1\le z\le N_1$}, 
\end{aligned}
\label{eq:ME1}
\end{equation}
and the transition rates in the states belonging to side branches $z\in\{0,N_1+1,N_1+2,\ldots N_2\}$ are given by
\begin{equation}
 \begin{aligned}
  R_{z\to z+1}&=R_{z+1\to z}=\frac{D}{\delta^2},\quad\text{for $N_1+1\le z\le N_2-1$},\\
   R_{0\to N_1+1}/2&=R_{N_1+1\to 0}=\frac{D}{\delta^2},
 \end{aligned}
 \label{eq:ME2}
\end{equation}
while all the remaining rates that are not listed in Eqs.~\eqref{eq:ME1} and \eqref{eq:ME2} are set to zero. Note that the division by 2 in the second line of Eq.~\eqref{eq:ME2} 
accounts for the degeneracy due to having two side-branches.
The generator of the master equation reads
\begin{equation}
 \mathcal{L}_{zz'}=
 \begin{cases}
  R_{z'\to z},&\text{if $z\neq z'$},\\
  -R_z,&\text{if $z= z'$},
 \end{cases}
\end{equation}
where $R_z\equiv \sum_{z'\neq z} R_{z\to z'}$ and $z\in\{0,1,\ldots,N_1+N_2\}$.
We numerically perform a complex eigendecomposition
of the generator $\mathcal{L}$
\begin{equation}
 \mathcal{L}=-\sum_{i=0}^{\hidewidth N_1+N_2-1\hidewidth}\lambda_i\ket{\psi_i^{\rm R}}\bra{\psi_i^{\rm L}},
 \label{eq:eigendecomposition}
\end{equation}
where $\lambda_k$ is the $k$th eigenvalue and $\psi_i^{\rm L}$ (or $\psi_i^{\rm L}$) are the corresponding left (or right) eigenvectors which according to Eq.~\eqref{eq:eigendecomposition} are normalized
$\bra{\psi_i^{\rm L}}\kket{\psi_j^{\rm R}}=\delta_{ij}$; note that $\lambda_0=0$. 
According to Eq.~(52) in Ref.~\cite{lapo20} the
variance of the occupation time in all the ring states
becomes
\begin{equation}
 \sigma_\tau^2(t)=2t\sum_{i=1}^{N_1+N_2-1} \frac{V_{0i}V_{i0}}{\lambda_i^\dagger}\bigg[1-\frac{1-\e^{\lambda_i^\dagger t}}{\lambda_i^\dagger t}\bigg],
 \label{eq:numerical}
\end{equation}
where $V_{ij}=\sum_{k=0}^{N_1}\bra{\psi_i^{\rm L}}\kket{k}\bra{k}\kket{\psi_j^{\rm R}}$ and $\lambda_i^\dagger $ is the complex conjugate of the eigenvalue $\lambda_i$ (see also Ref.~\cite{lapo18}). The lines in Fig.~4 in the main text are obtained from Eq.~\eqref{eq:numerical} with $D=\beta=1,l=3,L=10$, and $\delta =0.01$, i.e.,
$N_1=300$ and $N_2=1000$.

\subsubsection{Analytical long-time asymptotics}

We now provide the background and intuition about Eq.~(6) that addresses the long time limit and is adopted from Eq. (61) in Ref.~\cite{lapo20}.
The
long time dispersion is characterized by the first term in Eq.~\eqref{eq:numerical}, i.e.,
\begin{align}
 \mathcal{D}&=\lim_{t\to\infty }\frac{\sigma_\tau^2(t)}{t}=
 2\sum_{i=1}^{N_1+N_2-1} \frac{V_{0i}V_{i0}}{\lambda_i^\dagger}\nonumber\\
 &=2\int_0^\infty \dd t\Bigg[ \sum_{i=0}^{N_1+N_2-1} V_{0i}V_{i0}\e^{-\lambda_i^\dagger t}- V_{00}V_{00}\Bigg],
\end{align}
where in the second line we used $\int_0^\infty \e^{-\lambda t}=1/\lambda$ and $\lambda_0=0$.
Identifying $V_{00}=\phi_m$ and $\sum_{i=0}^{N_1+N_2-1} V_{0i}V_{i0}\e^{-\lambda_i^\dagger t}=P(m,t|m)\phi_m=P(m,t,m)$ yields the first line in Eq. (6) in the main text. Note the exact solution corresponds to the limit $N_1,N_2\to\infty$.

To obtain the second line of Eq.~(6) in the main text, it proves convenient to Laplace transform the time domain $t\to s$
such that any function $f(t)$ becomes $\tilde f(s)=\int_0^\infty\e^{-st}f(t)\dd t$. In this case the
Fokker-Planck equation can be conveniently solved analytically. Moreover,  the long time dispersion is then obtained from
\begin{align}
 \mathcal{D}&=2\int_0^\infty \dd t[P(m,t,m)-\phi_m^2]\nonumber\\
 &=2\lim_{s\to 0}\Big[\tilde P(m,s,m)-\frac{\phi_m^2}{s}\Big].
 \label{eq:D_lim}
\end{align}
Thus it suffices to determine the Laplace transform, $\tilde
P(m,s,m)$, of the joint probability density $P(m,t,m)$ to be in the mobile state and return to it again at time $t$.

We define the propagator  as 
$P_y(x,t|x_0)=P_y(x,t)$ with the initial condition $P_y(x,0)=\delta
(x-x_0)$ and $P_y(0,0)=0$ (we start in the mobile ring).
The return probability is obtained from integrating the propagator
over the mobile region $P(m,t,m)=\int_0^l\dd x\int_0^l\dd x_0 P_0(x,t|x_0)\phi_m/l$, where $\phi_m/l=P_0(x_0,\infty)$ is the stationary probability density within the ring.
The propagator satisfies the Fokker-Planck equation \eqref{eq:FPE_comb_time}, which after Laplace transformation 
and setting $D\equiv\beta\equiv 1$
becomes 
\begin{equation}
 \begin{aligned}
  [F\del_x-\del_x^2+s]\tilde P_0(x,s|x_0)&=\delta(x-x_0)\\
  [-\del_y^2+s] \tilde P_y(0,s|x_0)=0
 \end{aligned}
 \label{eq:FPE_comb}
\end{equation}
where $0\le y\le L$ and $0\le x\le l$ and
with boundary conditions translating into $\tilde P_0(0_+,s|x_0)=\tilde P_0(l-0_+,s|x_0)=2\tilde P_{0_+}(0,s|x_0)$
and $\del_x \tilde P_0(0_+,s|x_0)+\del_y\tilde P_{0_+}(0,s|x_0)=
\del_x\tilde P_0(l-0_+,s|x_0)$ as well as $\del_y\tilde P_L(0,s|x_0)=0$.

The solution of Eq.~\eqref{eq:FPE_comb} for any $s$ can be obtained
straightforwardly using the ansatz $\tilde
P_y(0,s|x_0)=\psi_0(y)=a_0\cosh[\sqrt{s} (L-y)]$, which solves the
second  line of Eq.~\eqref{eq:FPE_comb} with boundary condition
 $\del_y\tilde P_L(0,s|x_0)=\psi_0'(L)=0$. Moreover, the first line of
Eq.~\eqref{eq:FPE_comb} is solved by 
\begin{equation}
\tilde P_{0}(x,s|x_0)=
 \begin{cases}
  \psi_1(x)&\text{if $x\le x_0$},\\
  \psi_2(x)&\text{if $x> x_0$},
 \end{cases}
\end{equation}
where we use $\psi_i(x)=a^+_i\e^{\mu_+ x}+a^-_i\e^{\mu_- x}$ for
$i=1,2$ with $\mu_\pm=F/2\pm\sqrt{F^2/4+s}$. Note that the five parameters
$a_0,a^+_1,a^-_1,a^+_2,a^-_2$ are determined from the boundary conditions which read
\begin{equation}
 \begin{aligned}
  \psi_1(0)&=\psi_2(l)=\psi_0(0)/2,
  &\psi_1(x_0)&=\psi_2(x_0),\\
  \psi_2'(l)&=\psi_0'(0)+\psi_1'(0)
  &\psi_1'(x_0)-\psi_2'(x_0)&=1,
 \end{aligned}
 \label{eq:boundaries_Laplace}
\end{equation}
where the second condition of the last line follows from the
inhomogeneity caused by ``$\delta(x-x_0)$'' in
Eq.~\eqref{eq:FPE_comb}. The results of $a_0,a^+_1,a^-_1,a^+_2,a^-_2$
are too lengthy to be displayed and it turned out that they do not need to be precisely known.

We first perform the integral over the mobile ring ``$x$''
\begin{multline}
 \tilde P(m,s|x_0)\equiv\int_0^l\dd x
\tilde  P_{0}(x,s|x_0)
 \\
 =\frac{1}{s}-\int_0^L\dd y\tilde P_{y}(0,s|x_0)=\frac{1}{s}-a_0\frac{\sinh(\sqrt{s}L)}{\sqrt{s}},
 \label{eq:almost_return}
\end{multline}
where in the first step in the second line we used that the probability is conserved and  a constant ``1'' after Laplace transform in time becomes ``$1/s$'', while in the very last step we identified the parameter $a_0$, which does not depend on $x$ whereas it does depend on $x_0$ and $s$.
The joint return probability is obtained from Eq.~\eqref{eq:almost_return} via
\begin{align}
\tilde P(m,s;m)&=\int_0^l\dd x_0\tilde P(m,s|x_0)\frac{1}{2L+l}
\nonumber
\\
&=\frac{\phi_m}{s}-\frac{\sinh(\sqrt{s}L)}{\sqrt{s}(2L+l)}\int_0^la_0\dd x_0
\label{eq:prob_return}
\end{align}
Inserting  $a_0$, which solves the system of equations Eq.~\eqref{eq:boundaries_Laplace}, into Eq.~\eqref{eq:prob_return} and using Eq.~\eqref{eq:D_lim}  finally yields
\begin{align}
\mathcal{D}&=2\lim_{s \to 0}\Bigg[\frac{\phi_m(1-\phi_m)}{s}+\frac{\sinh(\sqrt{s}L)}{\sqrt{s}(2L+l)}\int_0^la_0\dd x_0\Bigg]
\nonumber\\
&=
\frac{4lL^2(F^2lL+3Fl\coth(Fl/2)-6)}{3F^2(l+2L)^3}
\label{eq:prob_return_fin}.
\end{align}
 In the last step we have used the computer algebra program
 \textsc{wolfram mathematica} which allowed us to conveniently carry out these straightforward albeit tedious calculations. 
This final step in \eqref{eq:prob_return_fin}  finally proves Eq.~(6) in the main text. Note we here used $\beta=D=1$. To restore the units we use $\mathcal{D}\to \mathcal{D}/D$ and $F\to\beta F$.

 \end{document}